\documentclass{aa}  
\def \eg {e.g.}

\def \ie {i.e.}
\def \cf {cf.}
\def \lcdm {{\hbox{$\Lambda$CDM}}}
\def \omegam {{\hbox{$\Omega_{\rm m}$}}}
\def \omegal {{\hbox{$\Omega_\Lambda$}}}
\def \hzero {{\hbox{$H_0$}}}
\def \arcmin {\hbox{$^\prime$}}
\def \arcsec {\hbox{$^{\prime\prime}$}}
\def \deg {\hbox{$^\circ$}}

\def \msun {\hbox{${\rm M_\odot}$}}

\def \rfive {\hbox{$r_{500}$}}
\def \mtwo {\hbox{$M_{200}$}}

\def \rtwo {\hbox{$r_{200}$}}

\newcommand{\kmsmpc }{\mbox{km s$^{-1}$ Mpc$^{-1}$}}

\newcommand{\mjyb }{\mbox{mJy beam$^{-1}$}}
\newcommand{\mujyb }{\mbox{$\mu$Jy beam$^{-1}$}}

\newcommand{\whz }{\mbox{W Hz$^{-1}$}}

\newcommand{\wsclean }{\textsc{WSClean}}

\newcommand{\pybdsf }{\textsc{pybdsf}}
\newcommand{\pybdsfE }{PYthon Blob Detector and Source Finder}

\newcommand{\aoflagger }{\textsc{AOFlagger}}
\newcommand{\dysco }{\textsc{Dysco}}
\newcommand{\dppp }{DP3}
\newcommand{\dpppE }{Default PreProcessing Pipeline}

\newcommand{\xmm }{{\em XMM-Newton}}

\newcommand{\xrism }{{\em XRISM}}

\newcommand{\ugmrt }{uGMRT}

\newcommand{\vla }{VLA}

\newcommand{\jvlaE }{Karl G. Jansky Very Large Array}
\newcommand{\lofar }{LOFAR}
\newcommand{\lofarE }{LOw Frequency ARray}

\newcommand{\meerkat }{MeerKAT}

\newcommand{\panstarrs }{Pan-STARRS}

\usepackage{graphicx}
\usepackage{amssymb}
\usepackage{multirow,bigdelim}
\usepackage{txfonts}
\usepackage[export]{adjustbox}
\usepackage{hyperref}
\hypersetup{colorlinks,citecolor=blue,filecolor=black,linkcolor=red,urlcolor=black}
\newcommand{\lband}{\textit{L}-band}
\newcommand{\oph}{Ophiuchus}
\newcommand{\ophcl}{Ophiuchus cluster}

\begin{document} 

\title{MeerKAT \textit{L}-band observations of the Ophiuchus galaxy cluster}
\subtitle{Detection of synchrotron threads and jellyfish galaxies} 

\authorrunning{Botteon et al.} 
\titlerunning{Synchrotron threads and jellyfish galaxies in the Ophiuchus cluster}

\author{Andrea Botteon\inst{\ref{ira}}, Marco Balboni\inst{\ref{unibo},\ref{iasf}}, Iacopo Bartalucci\inst{\ref{iasf}}, Fabio Gastaldello\inst{\ref{iasf}}, \and Reinout J. van Weeren\inst{\ref{leiden}}}

\institute{
INAF -- IRA, via P.~Gobetti 101, 40129 Bologna, Italy \label{ira} \\
\email{andrea.botteon@inaf.it} 
\and
Dipartimento di Fisica e Astronomia, Universit\`{a} di Bologna, via P.~Gobetti 93/2, 40129 Bologna, Italy \label{unibo}
\and
INAF -- IASF Milano, via A.~Corti 12, 20133 Milano, Italy \label{iasf}
\and
Leiden Observatory, Leiden University, PO Box 9513, 2300 RA Leiden, The Netherlands \label{leiden} \\
}

\date{Received XXX; accepted YYY}

\abstract
{Observations with modern radio interferometers are uncovering the intricate morphology of synchrotron sources in galaxy clusters, both those arising from the intracluster medium (ICM) and those associated with member galaxies. Moreover, in addition to the well-known radio tails from active galactic nuclei, radio continuum tails from jellyfish galaxies are being efficiently detected in nearby clusters and groups.}
{Our goal is to investigate the radio emission from the \ophcl, a massive, sloshing cluster in the local Universe ($z=0.0296$) that hosts a diffuse mini halo at its center.}
{To achieve this, we analyzed a 7.25 h \meerkat\ \lband\ observation, producing sensitive images at 1.28 GHz with multiple resolutions. A catalog of spectroscopically confirmed cluster galaxies was used to identify and study the member galaxies detected in radio.}
{We discover thin threads of synchrotron emission embedded in the mini halo, two of which may be connected to the brightest cluster galaxy. We also report the first identification of jellyfish galaxies in \oph, detecting six galaxies with radio continuum tails, one of which extending for $\sim$64 kpc at 1.28 GHz, making it one of the longest detected at such a high frequency. Finally, we propose an alternative scenario to explain the origin of a bright amorphous radio source, previously classified as a radio phoenix, aided by the comparison with recent simulations of radio jets undergoing kink instability.}
{
In \oph\ thin threads have been observed within the diffuse emission; a similar result was obtained in Perseus, another nearby cluster hosting a mini halo, suggesting that these structures may be a common feature in this kind of sources. Moreover, radio continuum observations have proven effective in detecting the first jellyfish galaxies in both systems.}



\maketitle
%

\section{Introduction}

The \ophcl\ is an intriguing target in the local Universe \citep[$z=0.0296$;][]{durret15}. It is a prime target for X-ray studies of the intracluster medium (ICM), being the second brightest galaxy cluster in the 2$-$10 keV band after Perseus \citep{edge90}. It hosts a cool core exhibiting strong temperature and metallicity gradients within its central 30 kpc, and a series of cold fronts indicative of gas sloshing \citep{ascasibar06, million10, werner16ophiuchus, gatuzz23ophiuchusvel}. It is a hot system \citep[$kT \sim 10$ keV;][]{matsuzawa96} and has been the subject of multiple studies on nonthermal X-ray emission \citep{eckert08, pereztorres09, nevalainen09, krivonos22}. Evidence of nonthermal electrons in the ICM is demonstrated by radio observations, which revealed a radio mini halo in the cluster center \citep{govoni09, murgia10ophiuchus}, \ie\ a kind of diffuse synchrotron emission that is generally observed in sloshing clusters \citep[\eg][]{giacintucci17}. A giant fossil radio lobe due to a past outburst of the active galactic nucleus (AGN) in the center of the cluster was also observed \citep{giacintucci20, giacintucci24proc2, giacintucci25sub}. \\
\indent
Despite the significant findings in X-ray and radio, the optical characterization of the \ophcl\ is challenged by its location at low Galactic latitude ($b = 9.3\deg$), where high dust extinction complicates the detection of cluster galaxies. Wide-field galaxy surveys suggest that \oph\ is embedded in a complex environment with numerous infalling substructures and is the most massive system within a supercluster \citep{hasegawa00, wakamatsu05proc}. From the analysis of 152 cluster galaxies with spectroscopic redshift, \citet{durret15} derived a large velocity dispersion, implying that \oph\ is a very massive system (see Table~\ref{tab:properties}), as expected from its high temperature and X-ray luminosity. \citet{durret15} also found that \oph\ galaxies have low star formation rates, detecting only two \textsc{H$\alpha$} emitters in the cluster. Among the candidate galaxy members identified through photometry, spiral galaxies appear indeed subdominant \citep{galdeano22}. The interpretation is that star formation is strongly suppressed because the \ophcl\ is a very massive and very old system, therefore environmental effects on member galaxies are as strong as can be in the Universe. \\
\indent
Observations of nearby galaxy clusters with modern interferometers, such as \meerkat\ \citep{jonas09} and \lofarE\ \citep[\lofar;][]{vanhaarlem13}, are providing unprecedented details on the diffuse emission arising from the ICM and that originating from cluster galaxies. For instance, \lofar\ observations have revealed the wealth of substructures in the mini halo occupying the cool core region of the Perseus cluster \citep{vanweeren24}, while both instruments are efficiently detecting thin (a few kpc in width) emission threads and ram pressure tails from cluster AGN \citep[\eg][]{ramatsoku20eso137, rudnick22, giacintucci22, velovic23} and star-forming galaxies \citep[\eg][]{roberts21a, deb22, ignesti23a2255, edler24}. The motivation of this work is to explore whether similar features are present in the \ophcl, which is an ideal target for such a study. Indeed, \oph\ is the nearest massive and sloshing cluster that \meerkat\ can observe, allowing us to search for emission threads both in the diffuse emission and in tailed AGN. Moreover, as radio observations are not affected by Galactic absorption, it is possible to use \meerkat\ continuum data to identify the most dramatic cases of stripped star-forming galaxies (\ie\ jellyfish galaxies) in the cluster, overcoming the difficulties of optical studies. \\
\indent
This paper is structured as follows. In Section~\ref{sec:reduction} we describe the reduction of the \meerkat\ \lband\ (856–1712 MHz) observations used. In Section~\ref{sec:results} we present the features and galaxies detected in our radio images. In Section~\ref{sec:discussion} we discuss our results in a broader context, while in Section~\ref{sec:conclusions} we summarize our conclusions. \\
\indent
In this work, we adopted a \lcdm\ cosmology with $\omegal = 0.7$, $\omegam = 0.3$ and $\hzero = 70$ \kmsmpc. At the cluster redshift ($z=0.0296$), this corresponds to a scale conversion factor of 0.593 kpc/arcsec (or, alternatively, 0.468 deg/Mpc).

\begin{table}
 \centering
 \caption{Properties of the \ophcl\ from \citet{durret15}.}\label{tab:properties}
 \begin{tabular}{cccc} 
  \hline
  \hline
  $z$ & $\sigma$ & \mtwo & \rtwo \\
   & [km s$^{-1}$] & [\msun] & [kpc] \\
  \hline
  0.0296 & 954 & $11.1\times10^{14}$ & 2125 \\
  \hline
 \end{tabular}
 \tablefoot{Reported quantities are redshift ($z$); galaxy velocity dispersion ($\sigma$); and mass within the radius inside which the mean mass density is 200 times the critical density at the cluster redshift (\mtwo, \rtwo).}
\end{table}

\section{Data reduction}\label{sec:reduction}

\begin{figure}
 \centering
 \includegraphics[width=\hsize,trim={0cm 0cm 0cm 0cm},clip]{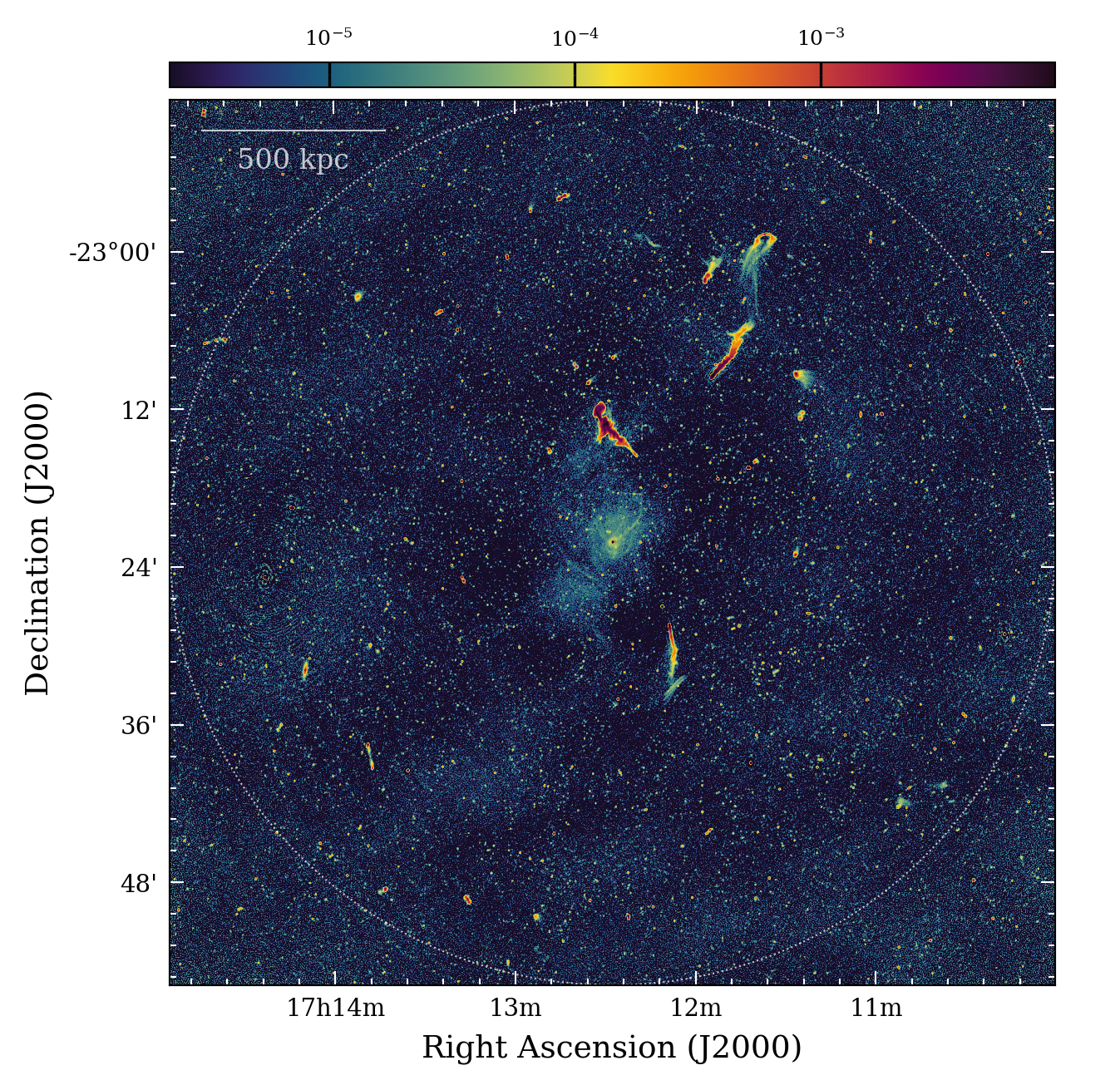}
 \caption{Wide-field \meerkat\ image with a robust $-0.5$ weighting of the \ophcl. The resolution and noise are $6.1\arcsec \times 5.5\arcsec$ and 4.5 \mujyb. The circle has a diameter of $67.3\arcmin$, corresponding to the FWHM $ = 57.5\arcmin \left( \frac{\nu}{1.5\rm{GHz}} \right)^{-1}$ of \meerkat\ computed at $\nu = 1.28$ GHz \citep{mauch20}. The color bar units are Jy beam$^{-1}$.}
 \label{fig:meerkat_fwhm}
\end{figure}

The \ophcl\ was observed with \meerkat\ \lband\ on 2020 March 21 for 7.25~h of on-source time (Capture Block ID: 1584831664). The observing strategy included 10~min scans of the flux and bandpass calibrators, either J1331+3030 (3C286) or J1939-6342, at hourly cadence. The target scans were 15~min long and were interleaved by 2~min scans on the gain calibrator J1733-1304. The observation covered a bandwidth of 856 MHz, divided into 4096 channels with a resolution of 208.984 kHz each. The central frequency of the data and the resulting images is 1.28 GHz. \\
\indent
For the data processing, we followed the steps described in \citet{botteon24a754} and briefly outlined here. We first downloaded from the \meerkat\ archive the data calibrated through the SARAO Science Data Processor calibration pipeline, flagged the visibilities affected by radio frequency interference with \aoflagger\ \citep{offringa10, offringa12}, averaged the data with the \dpppE\ \citep[\dppp;][]{vandiepen18}, and compressed the measurement set with \dysco\ \citep{offringa16}. After these preliminary steps, we performed self-calibration using the \texttt{facetselfcal.py}\footnote{\url{https://github.com/rvweeren/lofar_facet_selfcal}} script, originally developed by \citet{vanweeren21} for \lofar\ observations. Given the large extent of the \ophcl\ and our goal to image the entire region covered by \meerkat, we performed a direction-dependent calibration over the full field-of-view of the observation \citep[\cf][]{botteon24a754}. Specifically, we used five directions in which three self-calibration rounds were performed. \\
\indent
Final imaging was carried out with \wsclean\ v3.5 \citep{offringa14} using the multiscale multifrequency deconvolution \citep{offringa17}. Our reference image, presented in Fig.~\ref{fig:meerkat_fwhm}, was produced with a robust $-0.5$ weighting of the visibilities \citep{briggs95}. Additionally, we generated a lower resolution image by applying a 15\arcsec\ Gaussian taper and a higher resolution image using a robust $-1.0$ weighting. To mitigate the contribution of large-scale Galactic emission, to which short baselines are particularly sensitive, we imaged baselines with $\rm{uv_{min}}>150\lambda$ (equivalent to an angular scale of 22.9\arcmin\ or a physical scale of 815 kpc at the cluster redshift). This threshold was selected after testing different $\rm{uv_{min}}$ values, and its impact on the study of the most extended emission from the cluster is discussed in Appendix~\ref{app:diffuse}. \\
\indent
Displayed images are corrected for the primary beam attenuation using the \texttt{EveryBeam}\footnote{\url{https://git.astron.nl/RD/EveryBeam}} library employed within \wsclean, unless stated otherwise. We assumed a uncertainty of 10\% on the absolute flux density scale.

\begin{figure*}
 \centering
 \includegraphics[width=\hsize,trim={0cm 0cm 0cm 0cm},clip]{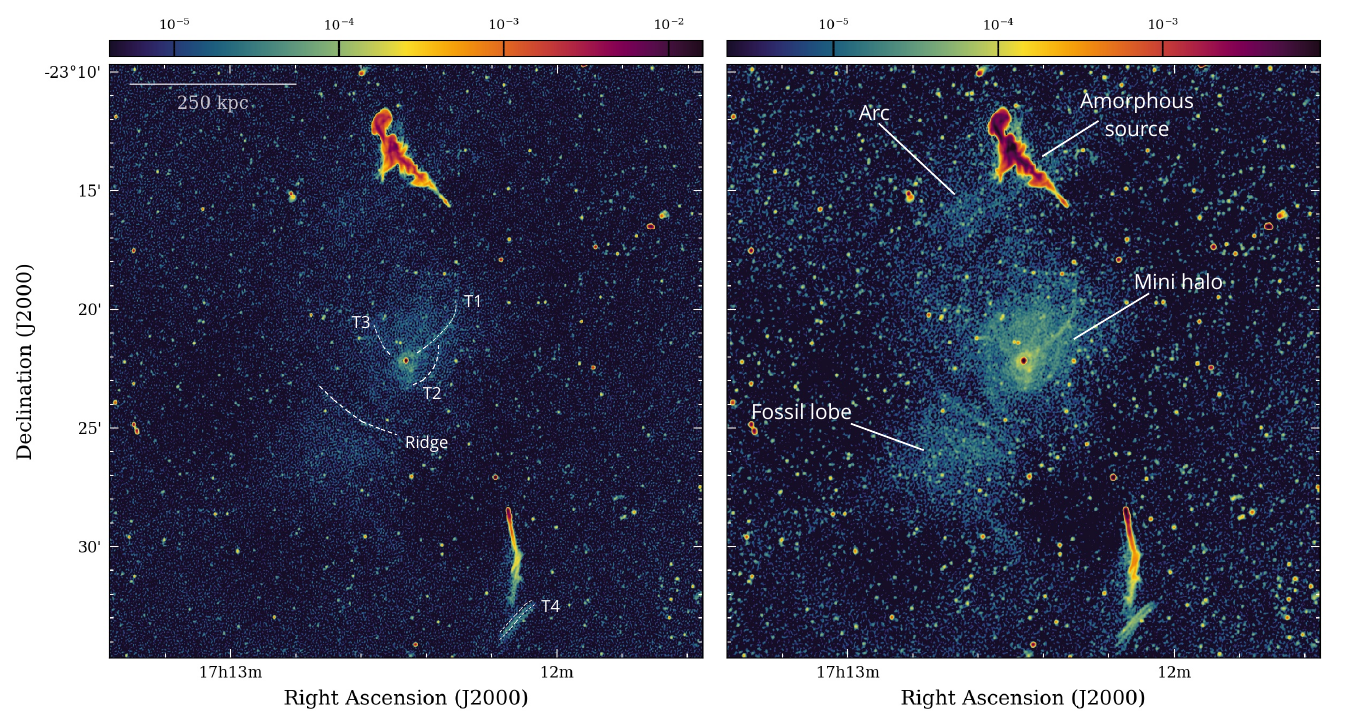}
 \caption{Central $25\arcmin \times 25\arcmin$ region of the \ophcl. \textit{Left:} \meerkat\ image with a robust $-1.0$ weighting with a resolution $4.8\arcsec \times 4.3\arcsec$ and noise 8.1 \mujyb. The dashed lines highlight the thready structures. \textit{Right:} zoom-in of the image shown in Fig.~\ref{fig:meerkat_fwhm} in which the diffuse sources are labeled. The color bar units are Jy beam$^{-1}$.}
 \label{fig:central}
\end{figure*}

\section{Results}\label{sec:results}

Our \meerkat\ wide-field image in Fig.~\ref{fig:meerkat_fwhm} shows the presence of numerous extended sources associated with both the ICM and cluster member galaxies. In the following sections, we discuss these two types of emission separately.

\subsection{Diffuse emission and filamentary structures}

\begin{figure*}
 \centering
 \includegraphics[width=\hsize,trim={0cm 0cm 0cm 0cm},clip]{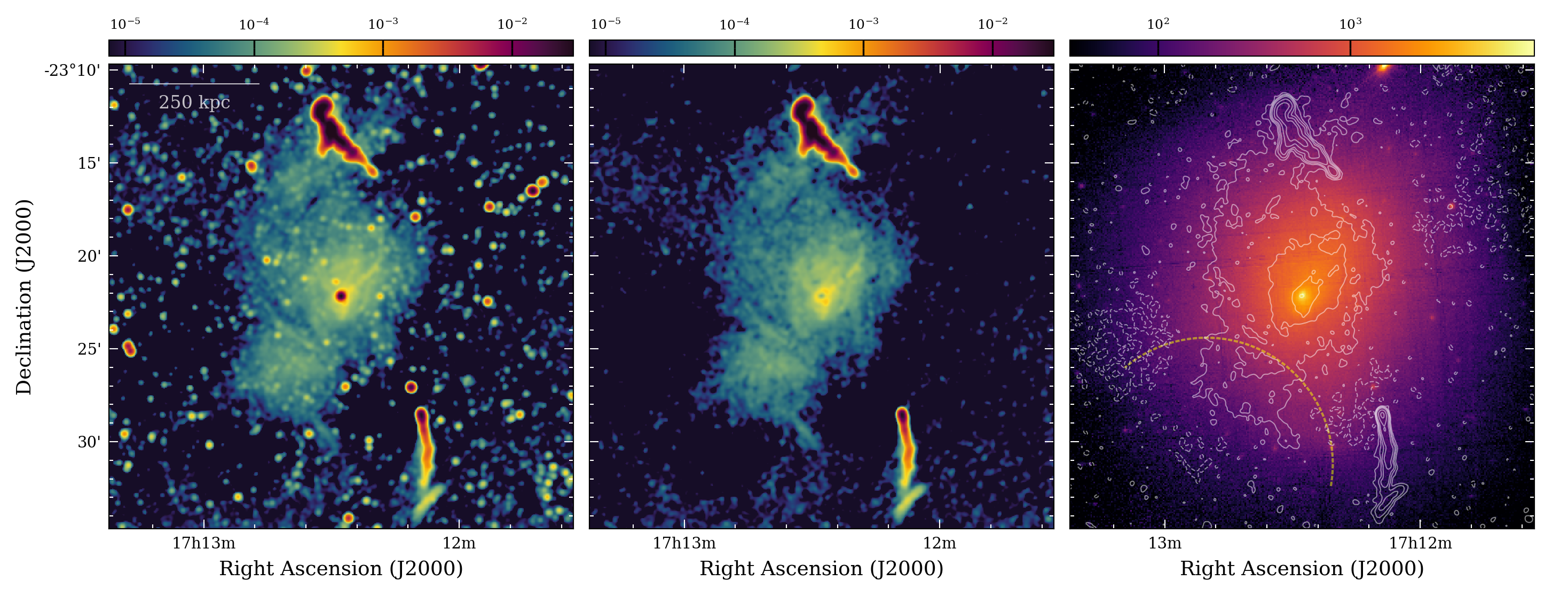}
 \caption{Radio diffuse emission and its connection with the thermal gas. The \textit{left} and \textit{central} panel show the \meerkat\ taper-15\arcsec\ images before and after source subtraction. The \textit{right} panel shows an \xmm\ wavelet cleaned image in the 0.5--2.5 keV band where the contours from the \meerkat\ image shown in the central panel are drawn at levels of [-0.03, 0.03, 0.10, 0.25, 0.80, 5, 20] \mjyb\ (the negative contour is reported in dashed). The yellow arc marks the X-ray concave edge. The color bar units are Jy beam$^{-1}$ for the radio images and counts for the X-ray image.}
 \label{fig:taper15_xmm}
\end{figure*}

In Fig.~\ref{fig:central} we report \meerkat\ images of the central region of \oph\ at two different resolutions. The most striking features revealed by our observations are three thin, elongated emission threads (or filaments) embedded within the diffuse emission. These structures are reported here for the first time and are labeled with the letter T in the left panel of Fig.~\ref{fig:central}. Among them, T1 stands out as the most prominent, with an increase in brightness up to $\sim$3 times with respect to the surrounding emission and extending at least 70 kpc in length (while maintaining a remarkably narrow width of less than 8 kpc). T1 and T3 seems to be connected with the brightest cluster galaxy (BCG) located at the center of the cluster and of the mini halo. A fourth similar structure, this time composed of two parallel threads and associated with a tailed radio galaxy, is also noted (T4). Another filamentary feature, likely corresponding to the ``ridge'' reported in \citet{giacintucci20}, is observed between the fossil lobe and mini halo. The largest linear size of the ridge is $\sim$150 kpc. \\
\indent
In addition to the newly discovered emission threads, it is worth noting that the diffuse emission from the mini halo and the fossil lobe southeast of the cluster core are clearly detected despite the relatively high resolution of the images shown. The mini halo extends northward and appears to be connected in projection to the amorphous source believed to be radio phoenix (labeled in Fig.~\ref{fig:central} and whose nature will be discussed in Section~\ref{sec:discussion_gal}). This ``bridge'' of emission was already claimed by \citet{giacintucci24proc2, giacintucci25sub} based on a \ugmrt\ image at 400 MHz with a beam of 18\arcsec. Our higher-resolution \meerkat\ images at 1.28 GHz not only confirm its presence but also reveal a more extended, arc-shaped structure. \\
\indent
With all due caveats discussed in Appendix~\ref{app:diffuse} regarding the analysis of the extended emission in \oph, in Fig.~\ref{fig:taper15_xmm} we report lower-resolution \meerkat\ images, both with and without discrete sources (see also Appendix~\ref{app:diffuse} for further details on the subtraction step), to enhance the detection of diffuse emission in the system. Based on these images, the largest linear size measured from the ridge to the arc is approximately 400--450 kpc, while the fossil lobe exhibits a roughly circular morphology with a radius of about 100 kpc (but \cf\ Appendix~\ref{app:diffuse}). The right panel of Fig.~\ref{fig:taper15_xmm} shows an overlay between the radio emission and the thermal gas as traced by \xmm\ archival observations. Notably, the ridge is co-located with the X-ray concave edge which marks the boundary of the giant cavity inflated by the fossil lobe \citep[see also][]{werner16ophiuchus, giacintucci20}.

\begin{figure*}
 \centering
 \includegraphics[width=\hsize,trim={0.3cm 8cm 10.7cm 0.3cm},clip]{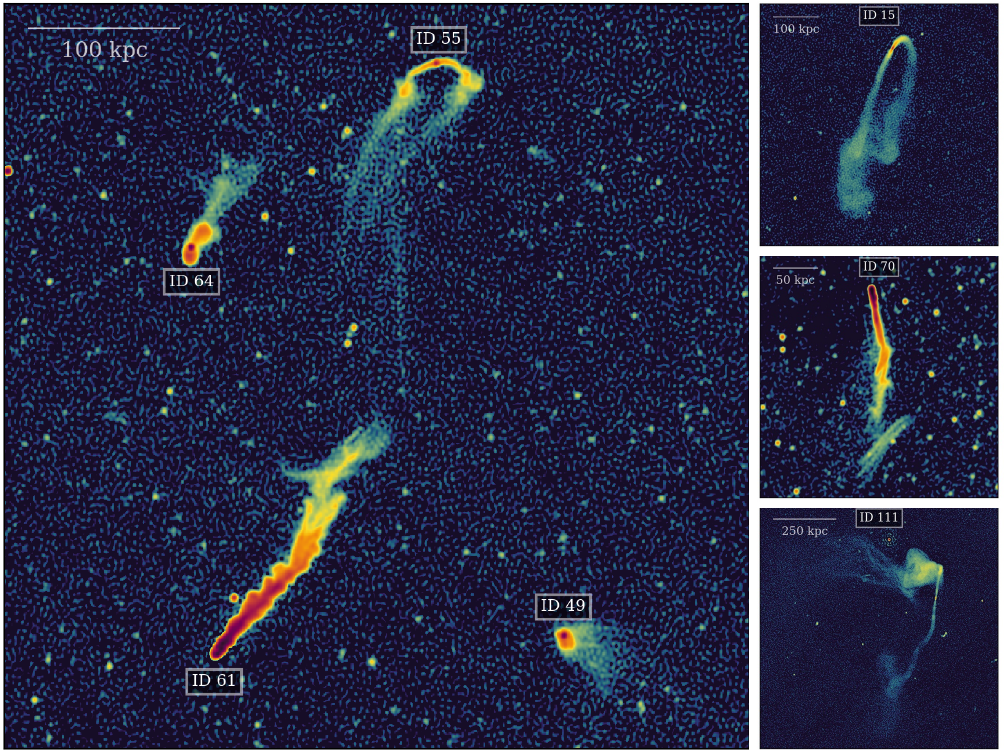}
 \caption{Different types of radio tails in \oph: four WATs (IDs 15, 55, 64, 111), two NATs (IDs 61, 70) and one jellyfish galaxy (ID 49). The large panel on the left displays the \meerkat\ image with a robust $-1.0$ weighting, while the smaller panels on the right show cutouts obtained from the image with a robust $-0.5$ weighting, not corrected for the primary beam response.}
 \label{fig:tail_collection}
\end{figure*}

\begin{figure}
 \centering
 \includegraphics[width=\hsize,trim={0cm 0cm 0cm 0cm},clip]{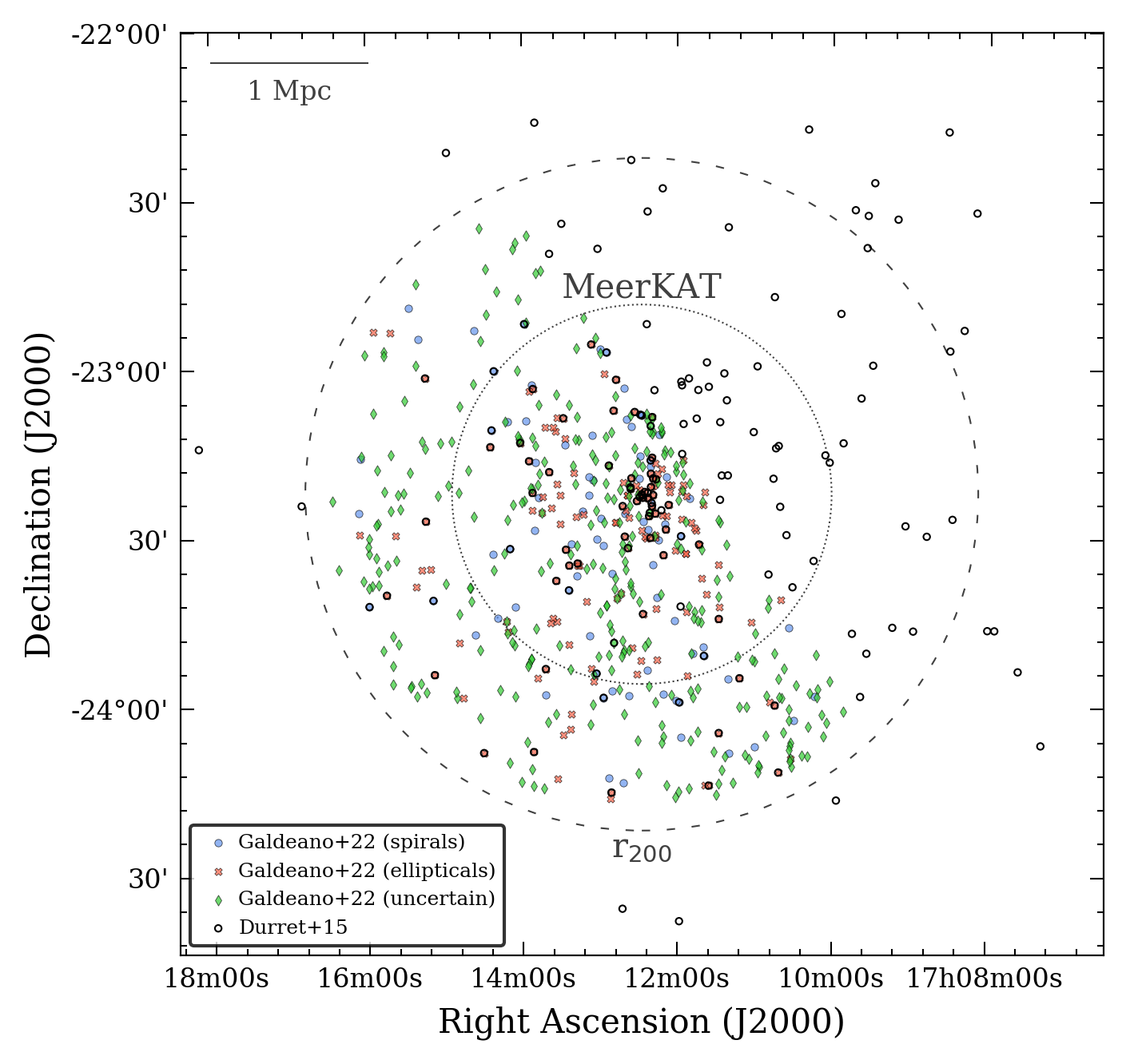}
 \caption{The 152 galaxies with spectroscopic redshift cataloged by \citet{durret15} and the 537 elliptical, spiral, and uncertain galaxies classified by \citet{galdeano22} in the \ophcl\ (see legend). The dotted and dashed circles denote the FWHM of \meerkat\ (as shown in Fig.~\ref{fig:meerkat_fwhm}) and \rtwo, respectively.}
 \label{fig:opt_cat}
\end{figure}

\begin{figure*}
 \centering
 \includegraphics[width=\hsize,trim={0cm 0cm 0cm 0cm},clip]{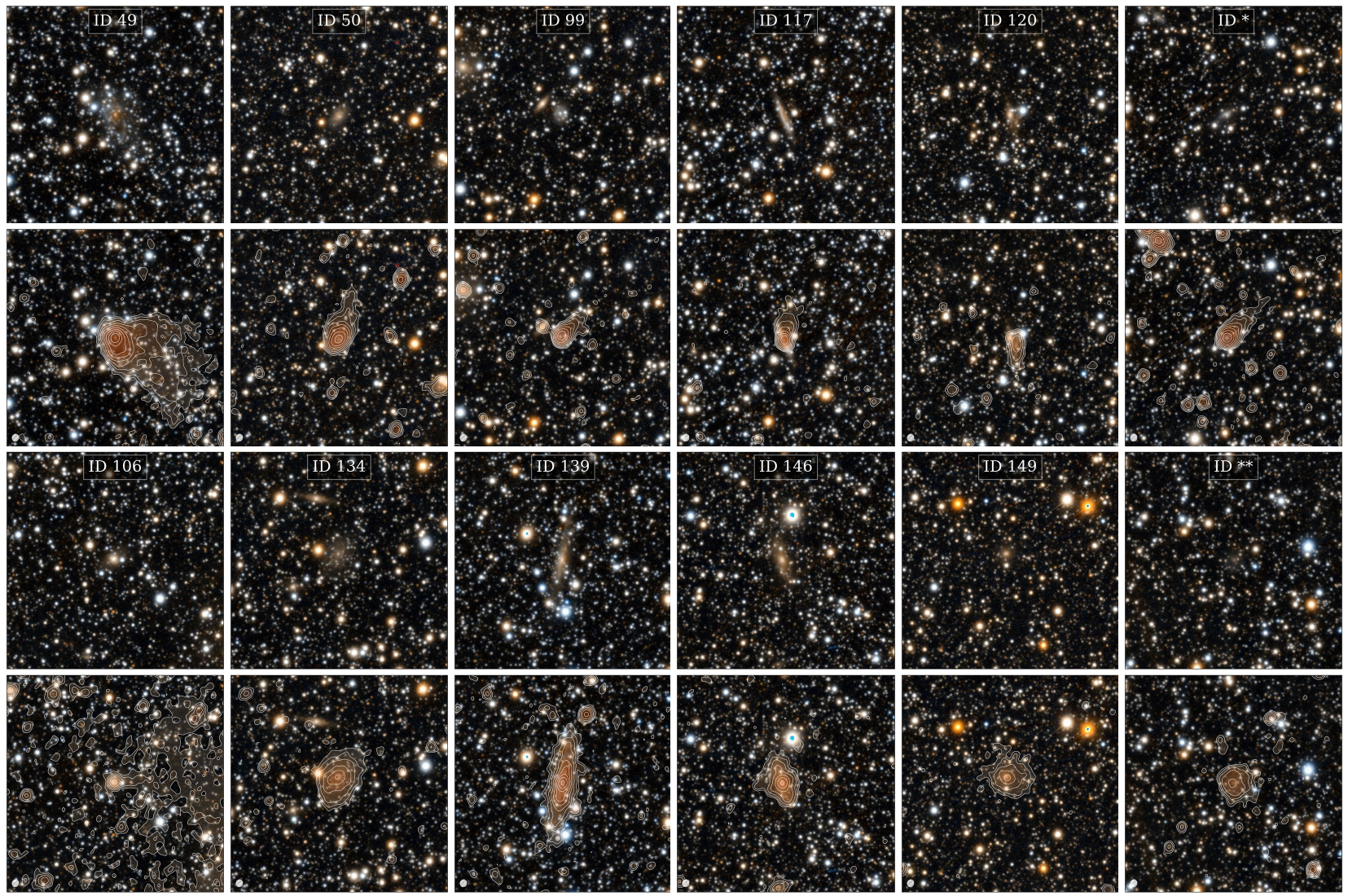}
 \caption{Star-forming galaxies in the \ophcl\ with detected radio emission. Each panel is 100 kpc $\times$ 100 kpc and shows the optical (\panstarrs, top) and optical/radio overlay (\panstarrs/\meerkat, bottom). The first six galaxies show pronounced tails and are classified as jellyfish galaxies. In contrast, the last six galaxies do not show clear evidence of radio tails, either due to the presence of extended emission from the disk or contamination from the mini halo (see \eg\ ID 106).}
 \label{fig:sfg}
\end{figure*}

\begin{figure}
 \centering
 \includegraphics[width=\hsize,trim={0cm 0cm 0cm 0cm},clip]{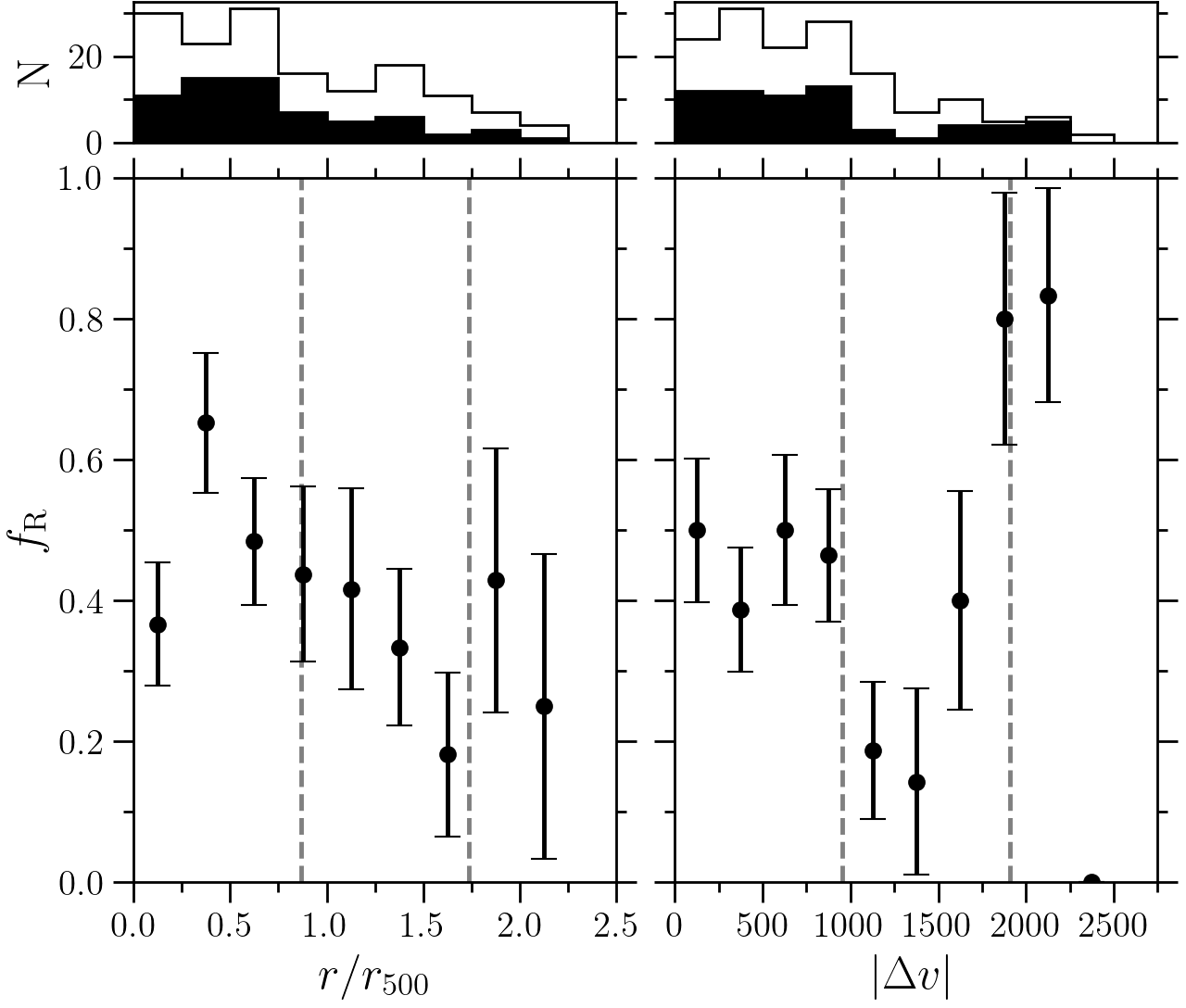}
 \caption{Fraction of spectroscopically confirmed cluster members that are detected in radio as a function of distance (\textit{left}) and velocity offset (\textit{right}, in km s$^{-1}$ units) from the BCG. Dashed vertical lines mark the position of $1\times$ and $2\times$ the \meerkat\ FWHM and the \oph\ galaxy velocity dispersion. The histograms show the distributions of the total number of spectroscopically confirmed cluster members (\textit{line}) and the subsample of those detected in radio (\textit{color}).}
 \label{fig:f_r}
\end{figure}

\subsection{Galaxies and tails}

\meerkat\ offers excellent sensitivity to both compact and extended sources, as demonstrated by the wide-field image presented in Fig.~\ref{fig:meerkat_fwhm}. As \oph\ is a low-redshift cluster, this is particularly useful to study the population of its radio-emitting member galaxies. Of particular interest are tailed radio sources, which form as a consequence of the motion of galaxies within the cluster environment. Tails can be either due to bended jets from AGN residing at the center of elliptical galaxies, or to ram-pressure stripped material from star-forming galaxies. The former produce bright tails that can extend up to megaparsec-scale and are generally classified as wide-angle tail (WAT) and narrow-angle tail (NAT) radio galaxies, depending on the bending angle between the jets \citep[\eg][]{miley80rev}. The latter, commonly referred to as jellyfish galaxies, generate fainter radio continuum tails that typically extend only a few tens of kiloparsecs from the stellar disk\footnote{Although an exceptional case of a 100 kpc ram-pressure tail was recently reported by \citet{roberts24ngc2276} at low frequency.} and are also observed at other wavelengths \citep[\eg][]{boselli22rev}. A visual inspection of our \meerkat\ images reveals that the \ophcl\ hosts all these types of tailed radio sources, as shown in the collection of selected extended radio galaxies presented in Fig.~\ref{fig:tail_collection}. \\
\indent
To confirm the membership of these galaxies to the \ophcl\ and identify other radio sources associated with cluster galaxies, we first produced a catalog of radio sources using the \pybdsfE\ \citep[\pybdsf;][]{mohan15}. Following \citet{hale25}, we adopted the default significance threshold (\texttt{thresh\_pix = 5}) and island boundary threshold (\texttt{thresh\_isl = 3}), set a zero background mean map (\texttt{mean\_map = `zero'}), and used a box size of 120 pixels with a step size of 30 pixels to compute the rms and mean over the entire image (\texttt{rms\_box = (120,30)}). As input image, we used the \meerkat\ image with a robust $-0.5$ weighting, not corrected for the primary beam response, which covers a region of $\sim$7.5 deg$^2$ (the image size is 8192 pixel $\times$ 8192 pixel, where 1 pixel = 1.2 arcsec), significantly larger than the area shown in Fig.~\ref{fig:meerkat_fwhm} (the actual coverage of this image is equal to the field shown in Fig.~\ref{fig:opt_cat}). The choice of using the uncorrected image was made to enable the detection of sources at large distances from the pointing center, where the polynomial approximation done by \wsclean\ for the primary beam is not longer valid. Two of such examples are the galaxies ID 15 and ID 111, shown in Fig.~\ref{fig:tail_collection}. \\
\indent
Once the radio catalog was generated, we cross-matched it with the catalog of 152 spectroscopically confirmed \oph\ cluster members published by \citet{durret15}. The positions of these galaxies, along with the candidate photometric members from \citet{galdeano22}, are shown in Fig.~\ref{fig:opt_cat}. We adopted a relatively large cross-match radius of 10\arcsec\ to account for potential offsets between the radio source coordinates provided by \pybdsf\ and the optical galaxy positions, which can occur when the radio emission is extended, such as in tailed sources. We note that in a few cases, such as galaxies ID 15, ID 55, and ID 61 shown in Fig.~\ref{fig:tail_collection}, the offset was even larger than 10\arcsec. Therefore, the cross-matched catalog and \meerkat\ image were carefully inspected visually, both to remove incorrect associations and to manually include missed extended sources. The final catalog comprises 65 galaxies and is presented in Table~\ref{tab:catalog}. When available, we also indicate their morphological classification as reported in \citet{galdeano22}. During the inspection, we identified three additional extended sources, likely associated with cluster members, that do not have a counterpart in \citet{durret15}. These sources are included at the bottom of the table. 
\begin{table}
 \centering
 \caption{Properties of the six jellyfish galaxies.}\label{tab:jellyfish}
 \begin{tabular}{lccc} 
  \hline
  \hline
  ID & Size & $S_{1.28}$ & $P_{1.28}$ \\
   & [kpc] & [mJy] & [\whz] \\
  \hline
  49 & 64 & $23.9\pm2.4$ & $(4.8\pm0.5)\times10^{22}$ \\
  50 & 34 & $7.1\pm0.7$ & $(1.4\pm0.1)\times10^{22}$ \\
  99 & 21 & $3.1\pm0.3$ & $(6.2\pm0.6)\times10^{21}$ \\
  117 & 26 & $3.2\pm0.3$ & $(6.4\pm0.6)\times10^{21}$ \\
  120 & 21 & $1.0\pm0.1$ & $(2.0\pm0.2)\times10^{21}$ \\
  * & 33 & $6.2\pm0.6$ & $(1.2\pm0.1)\times10^{22}$ \\
  \hline
 \end{tabular}
 \tablefoot{The projected largest linear size, flux density and power at 1.28 GHz are measured considering the emission above $3\sigma$ in the \meerkat\ image with a robust $-0.5$ weighting.}
\end{table}\\
\indent
As anticipated, the tailed sources shown in Fig.~\ref{fig:tail_collection} are confirmed cluster members. In general, the AGN-associated tails exhibit complex structures, characterized by bendings, wiggles, threads, and bars. For example, the NAT galaxy ID 70 extends for approximately 150 kpc before its structure appears to break, forming a bar-like feature inclined at $\sim$45\deg\ relative to the orientation of the tail. This bar consists of two parallel threads, labeled T4 in Fig.~\ref{fig:central}. This radio galaxy resembles the case of the broken-tail with a bar-like feature detected in SC J1329-313 \citep{venturi22}. A similar transverse feature is also seen in the final portion of the other NAT galaxy ID 61. Meanwhile, the WAT galaxies ID 55 and ID 111 display barbell-like morphologies, with long trails of relativistic plasma extending hundreds of kiloparsecs behind them, oriented orthogonally to the direction of the jets. Regarding star-forming galaxies, we identified at least six with tails extending beyond their stellar disks, which we therefore classified as jellyfish galaxies. These galaxies, that are the first jellyfish detected in the \ophcl, are shown in Fig.~\ref{fig:sfg}, along with six others that exhibit prominent radio emission. Noteworthy is the case of ID 49, whose tail extends for $\sim$64 kpc, making it one of the longest radio continuum tail detected in a jellyfish galaxy at frequencies above 1 GHz. In Fig.~\ref{fig:tail_collection}, the high-resolution \meerkat\ image reveals substructures within its tail that resemble tentacles. The main properties of the six jellyfish galaxies are summarized in Table~\ref{tab:jellyfish}. While \citet{roberts21a} found that in their sample of radio-selected jellyfish galaxies the presence of AGN activity is uncommon, multi-band data are necessary to assess potential AGN contamination and accurately determine the star-formation rates in the jellyfish of \oph.

\section{Discussion}\label{sec:discussion}

\subsection{Diffuse emission and filamentary structures}

\oph\ is the second closest galaxy cluster known to host a mini halo after Perseus. It is also the nearest mini halo cluster observable with \meerkat, as Perseus lies above the declination limit of the radio telescope. That is, the detail of our radio images offers a unique opportunity to study small-scale structures within the diffuse emission (while the analysis of the extended emission ``at large'' is challenged by the complexity of the field, see Appendix~\ref{app:diffuse}). This indeed allowed us to report the detection of thin synchrotron threads embedded in the mini halo. \\
\indent
In a recent work on the mini halo of Perseus, \citet{vanweeren24} highlighted the multitude of edges, filaments, and spurs using \lofar\ observations at 120--168 MHz, some of which can also be recognized in earlier \jvlaE\ (\vla) 270--430 MHz observations \citep{gendronmarsolais17}, although with lower resolution and sensitivity. The detection of similar features in the two closest known mini halos suggests that such structures are likely common in this class of sources and in radio halos more generally, as also probed by \lofar\ observations of the nearby merging clusters Abell 2255 \citep{botteon20a2255, botteon22a2255} and Coma \citep{bonafede22}. A key difference between (mini) halos resides in the dynamical state of their host clusters: mini halos are found in relaxed systems, whereas radio halos are associated with dynamically active, merging clusters \citep[\eg][]{vanweeren19rev}. In both cases, the origin of the radio-emitting particles is thought to be related to turbulent particle reacceleration in the ICM, with mini halos driven by sloshing-induced turbulence and radio halos powered by merger-induced turbulence \citep[\eg][]{brunetti14rev}. The different levels of turbulent velocities in these two types of systems are currently being investigated through \xrism\ observations \citep{xrism25a2029, xrism25centaurus}, which will soon target also \oph. \\
\indent
Sloshing motions result from the gravitational perturbation of the cluster core due to encounters with small infalling groups or subclusters. These interactions lead to a decoupling between dark matter and gas, causing the gas to ``slosh'' within the cluster potential well \citep[\eg][]{ascasibar06}. As a consequence of sloshing, concentric cold fronts and velocity shears are expected, and indeed have been reported in \oph\ \citep{million10, werner16ophiuchus, gatuzz23ophiuchusvel}. We note that the radio arc highlighted in Fig.~\ref{fig:central} (right panel) is located close to an area characterized by a transition from blueshifted to redshifted gas in the velocity spectral maps of \citet[][Fig. 13, bottom panel]{gatuzz23ophiuchusvel}. If this connection is real, the arc may be interpreted as nonthermal plasma that has been transported by shear flows, which in turn are expected to amplify the magnetic field strength \citep[see \eg][]{zuhone21bubbles}. A detailed study of the evolution of the ICM magnetic field in sloshing clusters was presented by \citet{zuhone11sloshing} using high-resolution magnetohydrodynamics (MHD) simulations, demonstrating how shear flows can stretch and amplify the field. This stretching leads to the formation of highly magnetized layers that follow the direction of the flow. Is this process related to the formation of the synchrotron threads detected in the cluster center? Their higher brightness compared to the surrounding medium provides indeed a possible indication that these features illuminate magnetic field structures. Alternatively, the threads could trace regions of enhanced cosmic ray electron density and/or reacceleration sites linked to the (past) nuclear activity of the BCG, to which T1 and T3 appear to be connected. If these structures are associated with the central galaxy, it is worth noting that the BCG in \oph\ does not show extended emission in the form of lobes or jets, and it is unresolved down to a resolution of $1.2\arcsec \times 2.5\arcsec$ (0.7 kpc $\times$ 1.4 kpc) \citep{werner16ophiuchus}, at odd with other radio galaxies where filamentary structures have been observed \citep[\eg][]{ramatsoku20eso137, condon21, brienza21, rudnick22, rajpurohit24ngc741}. For the same reason, it also differs from the BCG in Perseus. The mechanism responsible for formation of these threads remains unclear, highlighting the need for further investigation on the interplay between AGN activity, cosmic rays, magnetic fields, and gas motions in clusters.

\subsection{Galaxies and tails}\label{sec:discussion_gal}

As in all low-redshift clusters, extended radio galaxies with a broad variety of morphologies can be identified in \oph. But while the NAT and WAT galaxies in this system have been reported in previous studies \citep{govoni09, murgia10ophiuchus, werner16ophiuchus, giacintucci20, giacintucci24proc2, giacintucci25sub}, the jellyfish galaxies have been discovered only thanks to the radio continuum tails detected in the new \meerkat\ data presented here. As pointed out by \citet{roberts22perseus} in their discovery of four jellyfish galaxies in Perseus with \lofar, the identification of these sources in nearby clusters is particularly important, as it enables high-resolution multi-band studies (at $z \lesssim 0.05$, 1\arcsec\ corresponds to less than 1 kpc). In the \ophcl, which lies in the Zone of Avoidance, Galactic absorption hampers observations in other bands, making radio continuum observations especially valuable for discovering jellyfish galaxies, as demonstrated in this work. \\
\indent
With the data in hand, it is possible to investigate whether factors such as the distance to the cluster center and the galaxy velocity play a significant role in triggering the radio activity among the \oph\ cluster members. In the top panels of Fig.~\ref{fig:f_r}, we show the radial and velocity distributions of the spectroscopic cluster members \citep{durret15}, highlighting the subsample of those detected in radio (Table~\ref{tab:catalog}). The values are given in terms of distance from the BCG and absolute velocity offset relative to it. In the bottom panels, we present the fraction of members detected in radio, $f_{\rm R}$, within each radial and velocity bin, using $\Delta (r/\rfive) = 0.25$ and $\Delta( |\Delta v|) = 250$ km s$^{-1}$. Uncertainties on $f_{\rm R}$ are computed using binomial statistics, $\delta f_{\rm R} = \sqrt{[f_{\rm R} (1-f_{\rm R})]/N_{\rm tot}}$, where $N_{\rm tot}$ is the total number of galaxies in each bin. While no strong trends emerge from our data, it is worth noting that the peak values of $f_{\rm R}$ occur in the ranges $r/\rfive \in [0.25, 0.50]$ and $|\Delta v| \in [1750, 2250]$ km s$^{-1}$, \ie\ for galaxies that are relatively close to the cluster center and fast-moving. A word of caution is necessary regarding the limitations of our analysis. The radial distribution of $f_{\rm R}$ may be influenced by two factors. First, the presence of extended emission in the cluster center complicates the disentangling of galaxy emission embedded in the diffuse emission, possibly impacting the measurement of $f_{\rm R}$ in the innermost radial bin. Second, our catalog of cluster radio sources was derived from an image without primary beam correction applied. While this approach allows us to detect radio sources even at large distances from the instrument FWHM (\cf\ dashed lines in Fig.~\ref{fig:f_r}), the reduced sensitivity at these distances may result in missing fainter sources, potentially affecting $f_{\rm R}$ at large $r/\rfive$. The impact of the lower sensitivity in the outskirts on the velocity distribution is more subtle, as galaxies at large distances can exhibit both high and low velocity offsets. The only solution to this issue would require observing \oph\ with a mosaic of \meerkat\ pointings to ensure uniform sensitivity across the field. More broadly, a deep spectroscopic survey of \oph, despite the challenges posed by its low Galactic latitude, would also be beneficial for this kind of analysis, helping to minimize possible selection effects in the population of identified cluster members.
\begin{figure}
 \centering
 \includegraphics[width=\hsize,trim={0cm 0cm 0cm 0cm},clip]{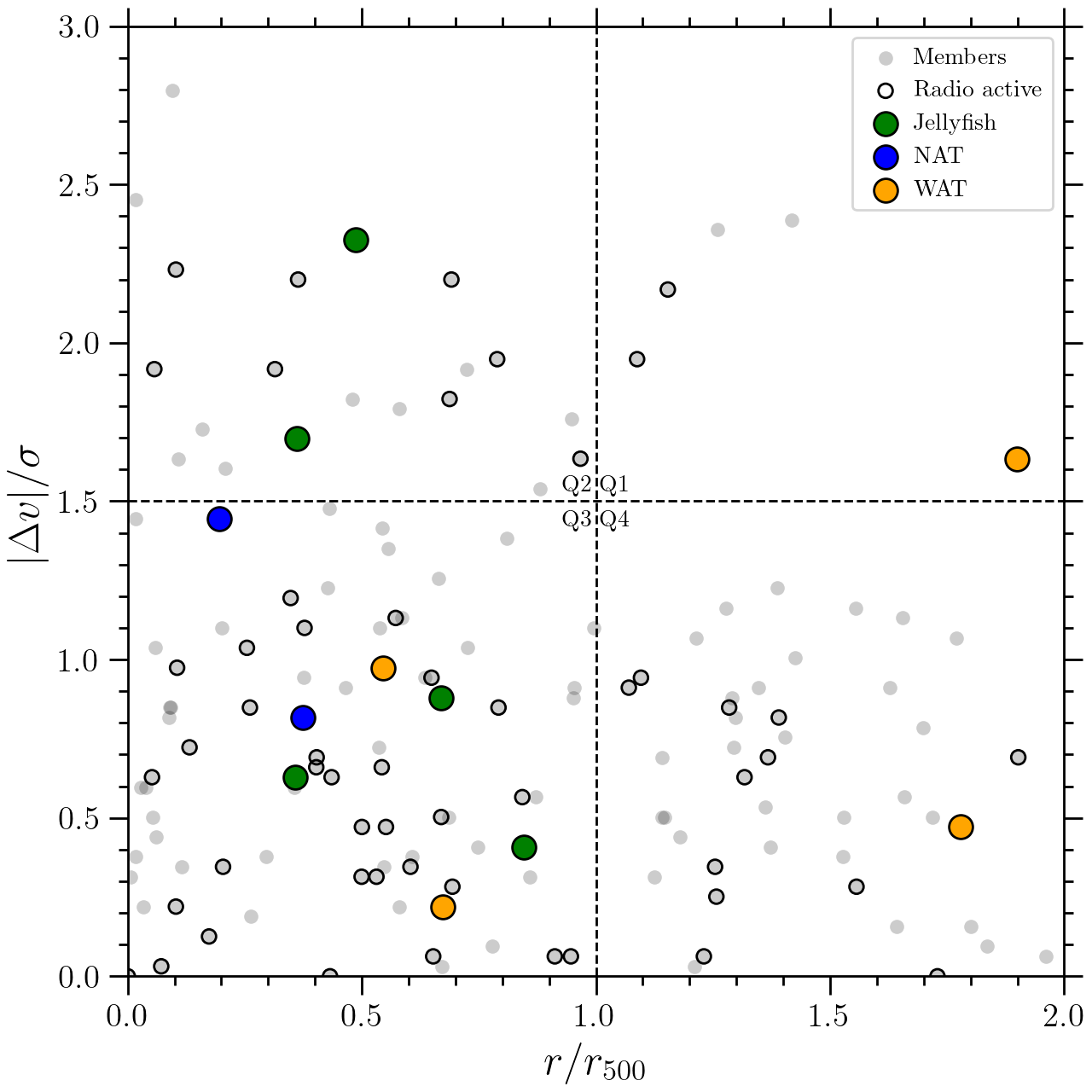}
 \caption{Projected phase-space of \oph\ member galaxies.}
 \label{fig:phase_space}
\end{figure} \\ 
\indent
Keeping the caveats described above in mind, we also constructed a projected phase-space diagram for the galaxies in \oph\ to examine $f_{\rm R}$ and the position of the radio tails in different regions of the diagram. This plot, shown in Fig.~\ref{fig:phase_space}, provides higher level information compared to the radial and velocity profiles of Fig.~\ref{fig:f_r} as it encloses the orbital history of cluster galaxies, which are expected to follow characteristic trajectories in phase-space during their infall into the cluster \citep[\eg][]{oman13, rhee17}. The velocity offset from the BCG is normalized to the cluster velocity dispersion to facilitate comparison with other clusters. Our plot spans the range $|\Delta v|/\sigma \in [0, 3]$ and $r/\rfive \in [0, 2]$ and is divided into four quadrants (Q1–Q4), defined by $|\Delta v|/\sigma = 1.5$ and $r/\rfive = 1$, following \citet{vanderjagt25sub}. We note that restricting the range to galaxies with $r/\rfive < 2$ excludes the four most distant spectroscopic cluster members, one of which is detected in radio, from the plot. In the four quadrants, the fractions of members detected in radio are $f_{\rm R}^{\rm Q1} = 0.60 \pm 0.22$, $f_{\rm R}^{\rm Q2} = 0.50 \pm 0.11$, $f_{\rm R}^{\rm Q3} = 0.47 \pm 0.06$, and $f_{\rm R}^{\rm Q4} = 0.30 \pm 0.07$. At large distances, significantly different fractions are found between the low-velocity region (Q4) and high-velocity region (Q1), although we note the low number statistics in Q1 and we remind the reader about the limitations of our analysis at large $r/\rfive$. Additionally, it is particularly interesting to pinpoint the position of the tailed radio sources in the phase-space diagram as they are expected to occupy specific regions due to the dependence of ram pressure on the ICM density and galaxy velocity ($\rho_{\rm ICM} v^2$). This relationship has been extensively investigated in previous studies on ram-pressure stripped galaxies in clusters \citep[\eg][]{mahajan11, haines15, jaffe18}, but only recently it has been applied specifically to radio continuum tails from jellyfish galaxies \citep{roberts21a, roberts21b} and AGN \citep{vanderjagt25sub}. The NAT, WAT, and jellyfish\footnote{The jellyfish galaxy ID * is not reported due to the lack of redshift.} galaxies in \oph\ are highlighted in Fig.~\ref{fig:phase_space} using different colors. All the detected jellyfish are located close to the cluster center (Q2-Q3). The lower sensitivity at larger distances may affect our ability to detect the faint tails of jellyfish galaxies in the outskirts. We note that \citet{roberts21a} found jellyfish to be in excess by factor of $\sim$2.5 compared to other star-forming cluster galaxies in Q1. We found that 4 out of the 6 considered NAT and WAT occupy region Q3, that is the region where all tailed radio galaxies in relaxed clusters reside in the sample of \citet{vanderjagt25sub}. However, we also identify two WAT galaxies at large cluster distances, while none was found in Q1 and only two, both in merging clusters, were identified in Q4 by \citet{vanderjagt25sub}. We note that tailed AGN in the sample of \citet{vanderjagt25sub} are at $z>0.1$, and resolution can play a role in the classification of the sources \citep{rudnick21tags}.
\begin{figure}
 \centering
 \includegraphics[width=\hsize,trim={0cm 0.7cm 0cm 0cm},clip]{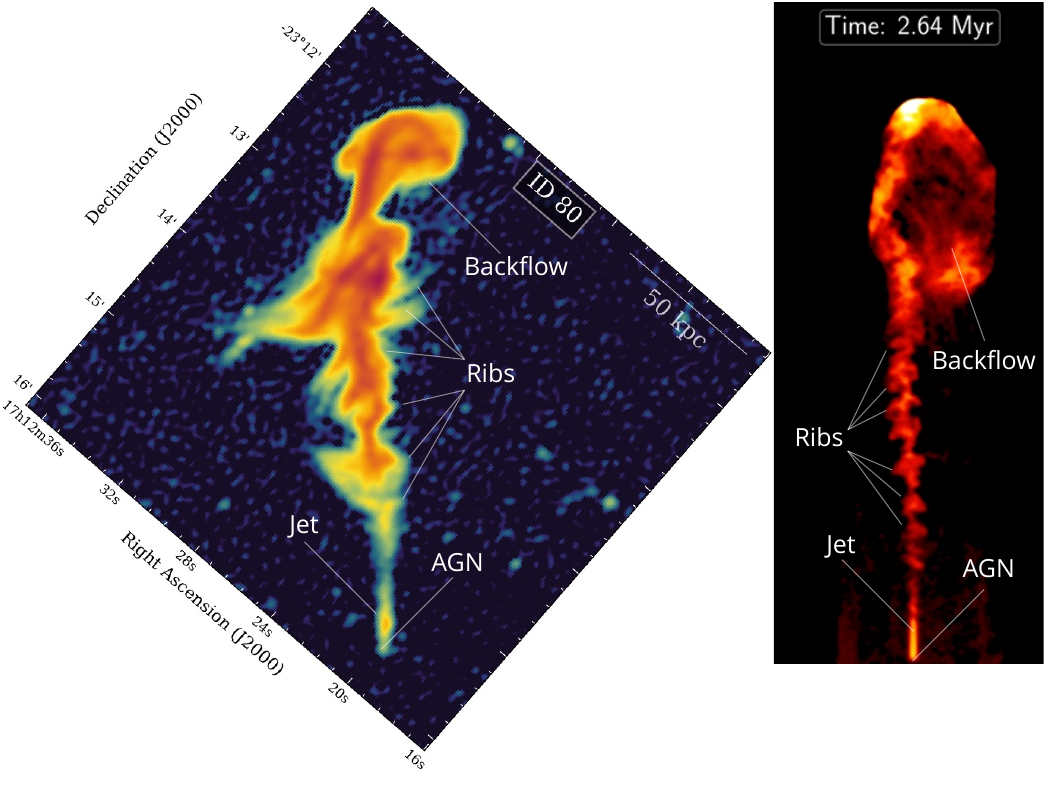}
 \caption{The amorphous radio source in \oph\ alongside a potential synthetic analog. \textit{Left:} \meerkat\ image with a robust $-1.0$ weighting rotated 41\deg\ clockwise to align the jet on the \textit{y}-axis. \textit{Right:} Snapshot from the numerical simulations of \citet{upreti24} showing the integrated synchrotron intensity at 1.28 GHz of a relativistic jet undergoing kink instability. The panel spans 50 kpc $\times$ 120 kpc. Similar features in the observations and simulations are labeled.}
 \label{fig:phoenix}
\end{figure} \\
\indent
We conclude this Section on the \oph\ cluster members detected in radio by discussing the case of the amorphous radio source, already labeled in Fig.~\ref{fig:central} (right panel). Its emission extends for $\sim$170 kpc and is highly filamentary, as shown in the zoom-in image reported in Fig.~\ref{fig:phoenix}. Due to its morphology, \citet{murgia10ophiuchus} proposed that it could be a radio phoenix, namely a class of steep-spectrum (\ie\ $\alpha > 1.5$, with $S_\nu \propto \nu^{-\alpha}$) synchrotron sources with filamentary morphologies that are typically interpreted as remnants of aged nonthermal plasma that have been revived by gas motions in the ICM \citep[\eg][]{vanweeren19rev}. However, the classification remained uncertain, as \citet{murgia10ophiuchus} found an integrated spectral index between 74 MHz and 1.5 GHz of $\alpha = 1.01\pm0.03$, too flat to be considered a radio phoenix. Instead, \citet{werner16ophiuchus}, using \vla\ A-array \lband\ data, reported in-band spectral indexes ranging from 1.8 to 2.5, leading them to conclude that the source is indeed likely a radio phoenix. In the following, we propose an alternative scenario to explain the nature of this source. \\
\indent
We associated galaxy ID 80, located at the south-west tip of the source, as progenitor of the emission. Along its extension, the source shows a number of transversal features, similarly to those of the ``mysterious'' tail (in short, MysTail) in Abell 3266 \citep{rudnick21a3266, riseley22a3266}, and those in the radio galaxies IC 711 in Abell 1314 \citep{vanweeren21} and ESO 137--G007 (nicknamed ``Corkscrew galaxy'') in Abell 3627 \citep{koribalski24corkscrew}. \citet{rudnick21a3266} referred to these structures as ``ribs'', and we adopt the same terminology hereafter. Another similarity between the amorphous source in \oph\ and MysTail is the absence of bright, compact emission at the position of the host galaxy, as well as a detachment between the host position and the rest of the emission. In a recent work, \citet{upreti24} used 3D relativistic MHD simulations to show that similar rib-like structures could emerge due to the wiggling and bending of a jet driven by kink instability. They compared their numerical results with the features observed in MysTail and found that a scenario in which jet activity is halted and then restarted better explains the distance of the ribs from the AGN and their separation with respect to a scenario involving continuous jet activity. They also found that deceleration near the jet head results in a backflow of jet material. We found a remarkable resemblance between the features in the amorphous source in \oph\ and the simulations with restarted jet activity of \citet{upreti24}, suggesting a common origin. In Fig.~\ref{fig:phoenix}, we present a direct comparison, highlighting the similarities between the observed emission and the synthetic radio emission. It is worth mentioning that the rib-like structures in the simulations and \meerkat\ image have comparable extensions of $\sim$10--20 kpc, with the exception of the two ribs farthest from the AGN, which extend for more than 50 kpc. As discussed in \citet{upreti24}, the jet in the MysTail may appear visible only on one side either because the source is genuinely one-sided or, in the case of a two-sided structure, the counter-jet is decelerated by the upwind or hindered by projection effects (\eg\ if it is strongly bent, making the structure appear one-sided); the same considerations can be applied to the amorphous source in \oph.


\section{Conclusions}\label{sec:conclusions}

We have used \meerkat\ \lband\ observations at 1.28 GHz to study the \ophcl, a massive system in the local Universe that is undergoing sloshing. In this paper:

\begin{enumerate}
    \item We discovered thin synchrotron threads embedded in the central diffuse emission. Their higher brightness with respect to the surrounding emission suggests that they may trace magnetic field structures or regions with higher cosmic ray electron densities or reacceleration sites.

    \item  We reported the first identification of jellyfish galaxies in the cluster based on the detection of their radio continuum tails.

    \item We investigated the fraction of cluster members detected in radio as a function of the distance from the cluster center, velocity offset from the BCG, and in the projected phase-space diagram.
    
    \item  We interpreted the features observed in the amorphous radio source, previously believed to be a radio phoenix, in the context of a radio galaxy whose jet is undergoing kink instability.
\end{enumerate}

Further work is needed to constrain the spectral index of the synchrotron threads and jellyfish tails, while a \meerkat\ mosaic would be required for a conclusive analysis of the population of radio sources in the cluster. A polarization analysis would instead be useful to determine whether the synchrotron threads trace coherent magnetic field structures and to obtain insights on the interaction between the tails with the surrounding medium. Distinctive polarization signatures are also predicted in the scenario of radio jets undergoing kink instability which we have adopted to interpret the amorphous radio source, with magnetic fields in the ribs primarily oriented perpendicular to the jet axis and less ordered in the backflow region \citep{upreti24}. \\
\indent
Recent \lofar\ observations of the Perseus cluster have led to the identification of the first jellyfish galaxies in the system \citep{roberts22perseus} and to the discovery of substructures embedded within its mini halo \citep{vanweeren24}. We obtained similar results for the \ophcl, highlighting the capability of highly sensitive radio observations to detect ram-pressure stripped galaxies and thin threads in the diffuse emission in nearby clusters. This suggests that such properties may be common in this type of systems.

\begin{acknowledgements}
We thank the anonymous referee for the comments aimed to improve the presentation of the manuscript.
We thank Stefan van der Jagt for discussion.
The \meerkat\ telescope is operated by the South African Radio Astronomy Observatory, which is a facility of the National Research Foundation, an agency of the Department of Science and Innovation.
This research made use of the LOFAR-IT computing infrastructure supported and operated by INAF, including the resources within the PLEIADI special ``LOFAR'' project by USC-C of INAF, and by the Physics Dept. of Turin University (under the agreement with Consorzio Interuniversitario per la Fisica Spaziale) at the C3S Supercomputing Centre, Italy.
This work made use of the following \textsc{python} packages: \texttt{APLpy} \citep{robitaille12}, \texttt{astropy} \citep{astropy22}, \texttt{CMasher} \citep{vandervelden20}, \texttt{EveryStamp}\footnote{\url{https://tikk3r.github.io/EveryStamp}}, \texttt{matplotlib} \citep{hunter07}, and \texttt{numpy} \citep{vanderwalt11}.
\end{acknowledgements}

%
%

\bibliographystyle{aa}
\bibliography{library.bib}

\begin{appendix}

\onecolumn

\section{Caveats on the analysis of the diffuse emission}\label{app:diffuse}

\begin{figure*}[b]
 \centering
 \includegraphics[width=\hsize,trim={0cm 0cm 0cm 0cm},clip]{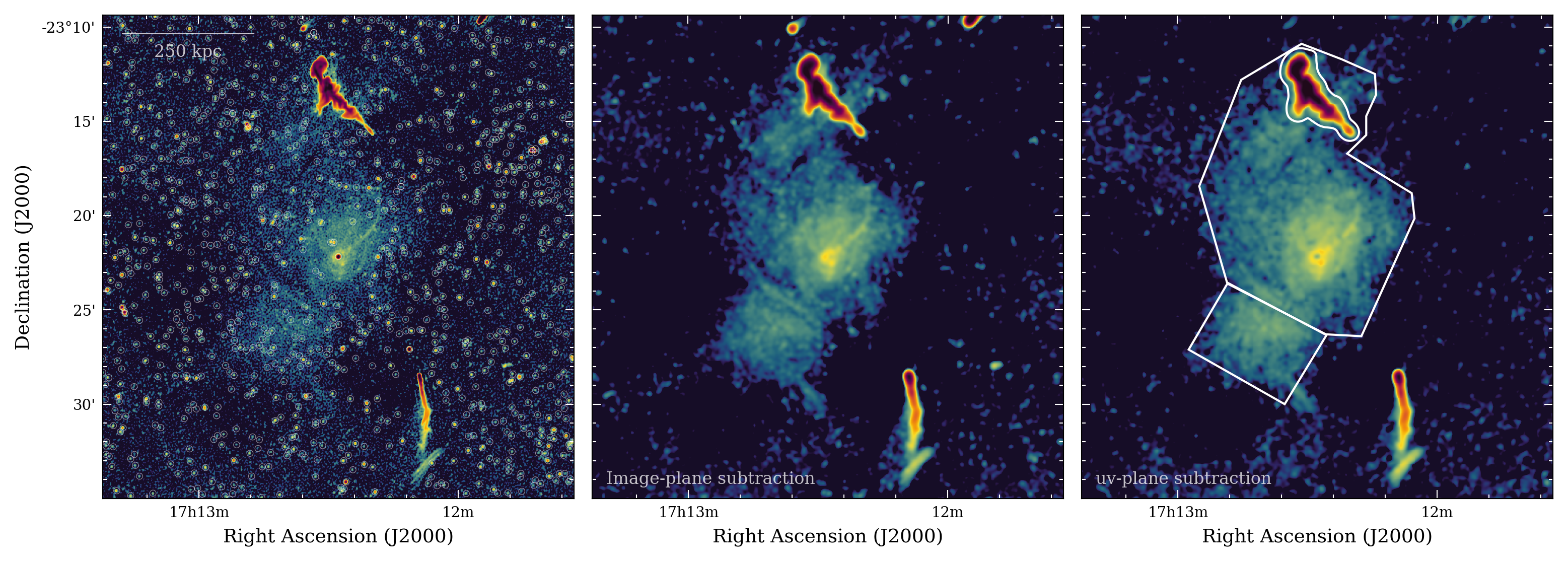}
 \caption{Different methods to subtract discrete sources (see text for more details).}
 \label{fig:sou_sub}
\end{figure*}

Discrete sources embedded in extended emission complicate the study of the diffuse component. Moreover, radio interferometers suffer from the lack of zero-spacing information, leading to the missing flux problem and the formation of negative bowls around extended sources. Both these effects can challenge the accurate measurements of the total flux density and morphology of the diffuse radio emission. In \oph, this type of analysis is further complicated by the cluster location at low Galactic latitude. In the following, we discuss these caveats in detail. \\
\indent
In Fig.~\ref{fig:sou_sub} we compare two different methods of source subtraction: one performed in the image-plane and the other in the uv-plane. 

\begin{itemize}
    \item In the image-plane subtraction, we started from the \meerkat\ image with a robust $-0.5$ weighting, where we identified the contaminating point sources and drew circles with radius of 10\arcsec\ around them (left panel). The pixel values within these regions were replaced with values extrapolated from the surroundings, as done in \citet{botteon23}. The resulting image was then convolved to a resolution of 15\arcsec\ to enhance the diffuse emission, producing the image shown in the central panel. The advantage of this method is that it is particularly fast, especially if a source detection algorithm is used to identify point sources, and computationally cheap. The disadvantage is that artifacts can be introduced when masking large areas (\eg\ in the case of large sources or closely clustered sources), and that by starting from a high-resolution image, the faint diffuse emission may not be fully deconvolved.

    \item In the uv-plane subtraction, a model consisting of the clean components of the contaminating sources was subtracted from the visibility data. The model was obtained from a high-resolution image, where the multiscale clean was constrained to use only two scales. This approach aims at minimizing the cleaning algorithm from picking up the diffuse emission in which sources are embedded while still capturing faint emission associated with discrete sources \citep[see][]{botteon24a754}. Before performing the subtraction, the model was carefully inspected to exclude clean components associated with the diffuse emission of interest and bright extended AGN, whose subtraction would be unreliable \citep{botteon22a2255, botteon24a754, vanweeren24}. The residual visibilities were then imaged using a Gaussian taper of 15\arcsec, producing the final image shown in the right panel (which is the same as the central panel in Fig.~\ref{fig:taper15_xmm}). The advantage of this method is that no artificial extrapolation of data is required and that the diffuse emission can be more easily recovered and deconvolved by performing imaging directly at low resolution. Its disadvantage is however the computational cost, as it requires two further imaging runs and the manipulation of the visibility data.
\end{itemize}

\begin{figure*}
 \centering
 \includegraphics[width=\hsize,trim={0cm 0cm 0cm 0cm},clip]{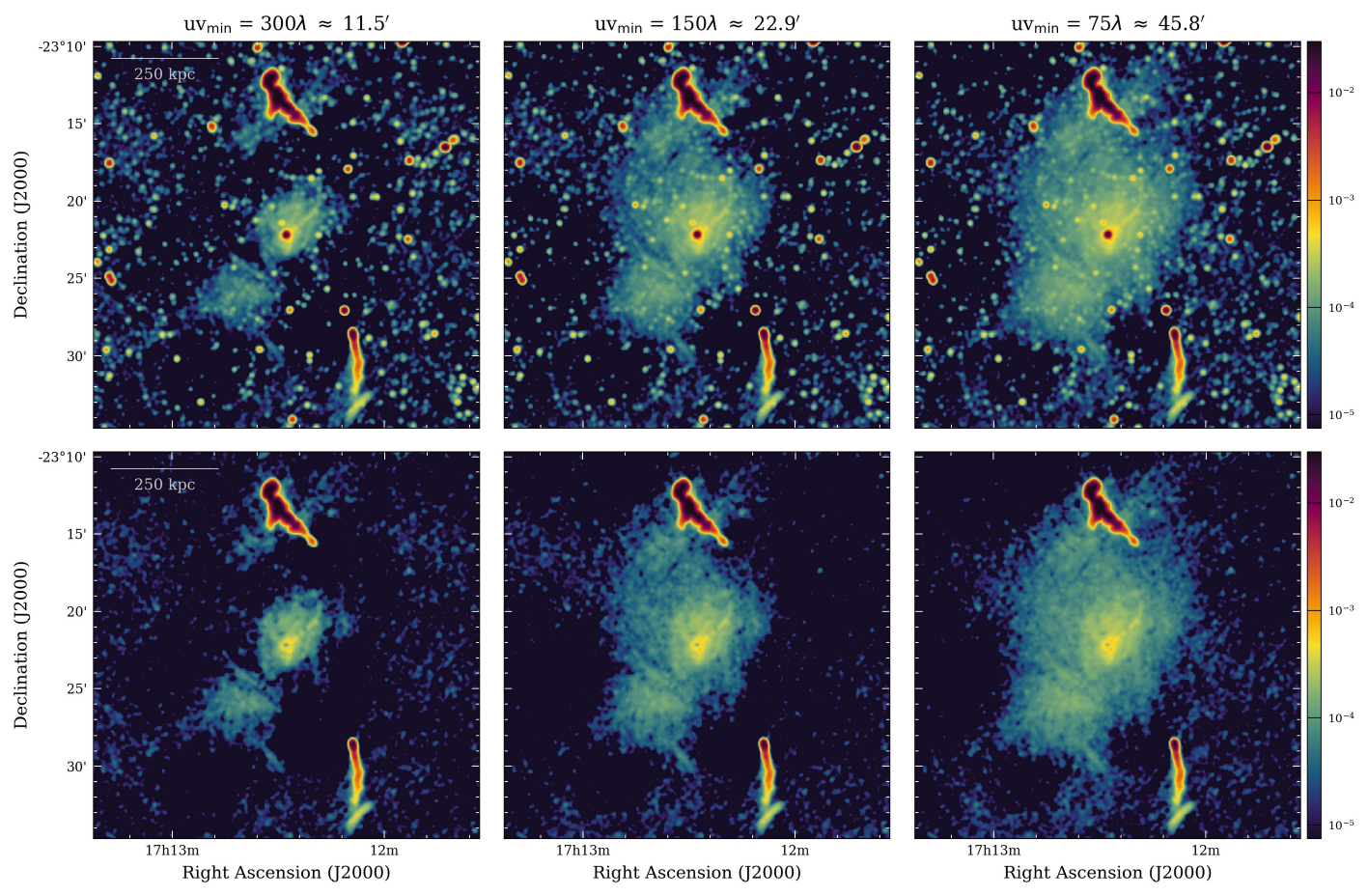}
 \caption{Impact of different uv$_{\rm min}$ values on the recovery of extended emission in \meerkat\ taper-$15\arcsec$ images with (\textit{top panels}) and without discrete sources (\textit{bottom panels}).}
 \label{fig:taper15_uvmin}
\end{figure*}

\noindent
We measured the flux densities of the fossil lobe and mini halo in both the image-plane (superscript ``img'') and uv-plane (superscript ``uv'') subtracted images by considering the same polygonal regions shown in the right panel of Fig.~\ref{fig:sou_sub}. For the fossil lobe, we measured $S_{1.28}^{\rm img} = 12.6\pm1.3$ mJy and $S_{1.28}^{\rm uv} = 17.1\pm1.7$ mJy. For the mini halo, we measured $S_{1.28}^{\rm img} = 69.2\pm6.9$ mJy and $S_{1.28}^{\rm uv} = 94.1\pm9.4$ mJy, where the flux density in the region encompassing the amorphous sources was extrapolated using the average surface brightness measured in the non-masked area of the mini halo. \\
\indent
The images discussed above, as well as those presented in the paper, were produced by imaging only baselines longer than $150\lambda$. In Fig.~\ref{fig:taper15_uvmin}, we show how using a uv$_{\rm min}$ value two times higher and lower affects the recovery of the extended emission. As is evident by eye, progressively excluding shorter baselines results in the loss of more diffuse emission. This is a well-known problem in interferometry and in the study of extended radio sources \citep[for an extensive discussion on this issue in \lofar, see][]{bruno23psz2dr2}. However, while including shorter baselines increases the sensitivity to the cluster diffuse emission, it also increases the sensitivity to the Galactic emission. This is particularly important when performing low-resolution imaging, as bright positive and negative patches of emission can appear across the entire image, affecting both flux density measurements and source morphology. This is the case of the \ophcl, which lays at low Galactic latitude, and where the need to include short baselines to recover most of the flux from the ICM is in competition with the the necessity to filter out large-scale Galactic emission. By considering the same polygons reported in Fig.~\ref{fig:sou_sub} (right panel), we measured flux densities of $S_{1.28}^{75\lambda} = 24.9\pm2.5$ mJy for the fossil lobe and $S_{1.28}^{75\lambda} = 130.2\pm13.0$ mJy for the mini halo in the uv$_{\rm min}$ = $75\lambda$ image. However, it is not possible to determine how much of the additional flux is due to genuine ICM emission recovered thanks to the lower uv$_{\rm min}$ and how much is contributed by unrelated foreground emission sitting on top of the cluster. Moreover, as the brightness of the emission patches can take absolute values\footnote{We remind the reader that the brightness can either be negative or positive.} that are more than 50\% the brightness of the peripheral region of the mini halo, the quoted errors on the flux densities are actually significant larger. This implies that, in general, the flux densities are poorly constrained. We therefore conclude that our \meerkat\ low-resolution images cannot be used for a quantitative analysis of the diffuse emission.

\newpage
\section{Member galaxies detected in radio}\label{app:radio_members}

\begin{table*}[h]
 \centering
 \caption{Ophiuchus radio galaxies.}\label{tab:catalog}
 \scriptsize 
 \begin{tabular}{llcccccccccc} 
  \hline
  \hline
  ID & Name & RA & DEC & $z$ & $r$ & $r/\rfive$ & $\Delta v$ & $|\Delta v|/\sigma$ & Mtype & Comment \\
  & & [deg] & [deg] & & [kpc] & & [km s$^{-1}$] & & & \\
  \hline
2 & 2MASX J17073583-2353321 & 256.8993 & -23.8923 & 0.0317 & 2627 & 1.90 & 660 & 0.69 &  &  \\
3 & 2MASX J17075426-2346173 & 256.9761 & -23.7715 & 0.0295 & 2389 & 1.73 & 0 & 0.00 &  &  \\
10 & 2MASX J17084712-2329350 & 257.1963 & -23.4931 & 0.0275 & 1820 & 1.32 & -600 & 0.63 &  &  \\
11 & 2MASX J17085724-2346260 & 257.2385 & -23.7739 & 0.0269 & 1921 & 1.39 & -779 & 0.82 &  &  \\
15 & 2MASX J17092831-2226505 & 257.3680 & -22.4474 & 0.0310 & 2457 & 1.78 & 450 & 0.47 &  & WAT \\
21 & 2MASX J17093810-2305055 & 257.4087 & -23.0849 & 0.0265 & 1514 & 1.10 & -899 & 0.94 &  &  \\
23 & 1RXS  J170944.9-234658 & 257.4362 & -23.7814 & 0.0364 & 1593 & 1.15 & 2069 & 2.17 &  &  \\
25 & 2MASX J17095391-2250048 & 257.4746 & -22.8347 & 0.0293 & 1700 & 1.23 & -60 & 0.06 &  &  \\
29 & 2MASX J17101464-2333587 & 257.5610 & -23.5663 & 0.0277 & 1164 & 0.84 & -540 & 0.57 &  &  \\
32 & 2MASX J17103573-2329248 & 257.6489 & -23.4902 & 0.0237 & 949 & 0.69 & -1739 & 1.82 &  &  \\
36 & 2MASX J17104398-2313598 & 257.6833 & -23.2333 & 0.0325 & 896 & 0.65 & 899 & 0.94 &  &  \\
38 & 2MASX J17104522-2247088 & 257.6884 & -22.7858 & 0.0357 & 1503 & 1.09 & 1859 & 1.95 &  &  \\
41 & 2MASX J17105842-2259266 & 257.7435 & -22.9907 & 0.0233 & 1090 & 0.79 & -1859 & 1.95 &  &  \\
44 & 2MASX J17112078-2234457 & 257.8366 & -22.5794 & 0.0322 & 1774 & 1.28 & 809 & 0.85 &  &  \\
45 & 2MASX J17112119-2318497 & 257.8383 & -23.3138 & 0.0274 & 556 & 0.40 & -630 & 0.66 &  &  \\
47 & 2MASX J17112403-2300427 & 257.8501 & -23.0119 & 0.0279 & 925 & 0.67 & -480 & 0.50 &  &  \\
49 & 6dF   J1711272-230922 & 257.8631 & -23.1562 & 0.0369 & 673 & 0.49 & 2218 & 2.33 &  & Jellyfish \\
50 & 2MASX J17112722-2323097 & 257.8634 & -23.3860 & 0.0315 & 495 & 0.36 & 600 & 0.63 &  & Jellyfish \\
53 & 2MASX J17113574-2413554 & 257.8990 & -24.2321 & 0.0317 & 1889 & 1.37 & 660 & 0.69 & 3 &  \\
55 & 2MASX J17113750-2258457 & 257.9062 & -22.9794 & 0.0288 & 929 & 0.67 & -210 & 0.22 &  & WAT \\
56 & OPH J171139.58-235053 & 257.9149 & -23.8481 & 0.0322 & 1094 & 0.79 & 809 & 0.85 & 5 &  \\
57 & 2MASX J17114344-2331057 & 257.9310 & -23.5182 & 0.0333 & 481 & 0.35 & 1139 & 1.19 & 3 &  \\
58 & 2MASX J17114409-2303397 & 257.9337 & -23.0610 & 0.0274 & 749 & 0.54 & -630 & 0.66 &  &  \\
60 & 2MASX J17115130-2301363 & 257.9638 & -23.0267 & 0.0331 & 791 & 0.57 & 1079 & 1.13 &  &  \\
61 & 2MASX J17115542-2309423 & 257.9810 & -23.1618 & 0.0269 & 517 & 0.37 & -779 & 0.82 &  & NAT \\
62 & 2MASX J17115651-2315013 & 257.9855 & -23.2504 & 0.0322 & 361 & 0.26 & 809 & 0.85 &  &  \\
63 & 2MASX J17115666-2302503 & 257.9861 & -23.0473 & 0.0305 & 734 & 0.53 & 300 & 0.31 &  &  \\
64 & 2MASX J17115724-2302113 & 257.9885 & -23.0365 & 0.0326 & 754 & 0.55 & 929 & 0.97 &  & WAT \\
68 & 2MASX J17115882-2359083 & 257.9951 & -23.9856 & 0.0347 & 1336 & 0.97 & 1559 & 1.63 & 5 &  \\
69 & 2MASX J17120697-2324053 & 258.0290 & -23.4015 & 0.0318 & 182 & 0.13 & 690 & 0.72 & 3 &  \\
70 & 2MASX J17120908-2328263 & 258.0378 & -23.4740 & 0.0249 & 269 & 0.20 & -1379 & 1.45 & 3 & NAT \\
78 & 2MASX J17121895-2322192 & 258.0790 & -23.3720 & 0.0275 & 72 & 0.05 & -600 & 0.63 & 3 &  \\
80 & 2MASX J17121976-2315418 & 258.0824 & -23.2616 & 0.0299 & 240 & 0.17 & 120 & 0.13 & 3 & Amorphous source \\
81 & 2MASX J17121989-2324187 & 258.0829 & -23.4052 & 0.0296 & 99 & 0.07 & 30 & 0.03 & 3 &  \\
84 & 2MASX J17122070-2318318 & 258.0863 & -23.3088 & 0.0366 & 142 & 0.10 & 2129 & 2.23 & 3 &  \\
85 & 2MASX J17122085-2310048 & 258.0869 & -23.1680 & 0.0356 & 434 & 0.31 & 1829 & 1.92 & 0 &  \\
87 & 2MASX J17122113-2329568 & 258.0880 & -23.4991 & 0.0306 & 281 & 0.20 & 330 & 0.35 & 3 &  \\
89 & 2MASX J17122214-2326048 & 258.0923 & -23.4347 & 0.0264 & 146 & 0.11 & -929 & 0.97 & 3 &  \\
98 & 2MASX J17122774-2322108 & 258.1156 & -23.3697 & 0.0295 & 0 & 0.00 & 0 & 0.00 & 3 & BCG \\
99 & 2MASX J17122782-2308108 & 258.1159 & -23.1363 & 0.0349 & 498 & 0.36 & 1619 & 1.70 & 5 & Jellyfish \\
100 & 2MASX J17122884-2308038 & 258.1202 & -23.1344 & 0.0365 & 503 & 0.36 & 2099 & 2.20 & 5 &  \\
103 & 2MASX J17123325-2307348 & 258.1386 & -23.1263 & 0.0330 & 522 & 0.38 & 1049 & 1.10 & 3 &  \\
106 & 2MASX J17123638-2321138 & 258.1516 & -23.3538 & 0.0356 & 79 & 0.06 & 1829 & 1.92 & 3 & Uncertain \\
108 & 2MASX J17123843-2331458 & 258.1601 & -23.5294 & 0.0262 & 352 & 0.25 & -989 & 1.04 & 3 &  \\
110 & 2MASX J17124256-2324167 & 258.1773 & -23.4047 & 0.0288 & 142 & 0.10 & -210 & 0.22 & 3 &  \\
111 & 2MASX J17124278-2435477 & 258.1783 & -24.5966 & 0.0243 & 2622 & 1.90 & -1559 & 1.63 &  & WAT \\
112 & OPH J171249.23-234834 & 258.2051 & -23.8095 & 0.0365 & 955 & 0.69 & 2099 & 2.20 & 0 &  \\
114 & 2MASX J17124957-2307207 & 258.2065 & -23.1224 & 0.0273 & 557 & 0.40 & -660 & 0.69 & 3 &  \\
117 & 2MASX J17125506-2257000 & 258.2294 & -22.9500 & 0.0323 & 924 & 0.67 & 839 & 0.88 & 5 & Jellyfish \\
120 & 2MASX J17130287-2354020 & 258.2620 & -23.9006 & 0.0282 & 1169 & 0.85 & -390 & 0.41 & 5 & Jellyfish \\
122 & 2MASX J17131749-2334290 & 258.3229 & -23.5747 & 0.0295 & 597 & 0.43 & 0 & 0.00 & 3 &  \\
124 & 2MASX J17132420-2339146 & 258.3509 & -23.6541 & 0.0280 & 762 & 0.55 & -450 & 0.47 & 5 &  \\
126 & 2MASX J17132844-2308406 & 258.3685 & -23.1446 & 0.0280 & 691 & 0.50 & -450 & 0.47 & 3 &  \\
130 & OPH J171339.33-231814 & 258.4139 & -23.3041 & 0.0315 & 601 & 0.44 & 600 & 0.63 & 3 &  \\
131 & 2MASX J17134226-2353126 & 258.4261 & -23.8868 & 0.0297 & 1260 & 0.91 & 60 & 0.06 & 3 &  \\
134 & 2MASX J17135195-2303286 & 258.4665 & -23.0579 & 0.0304 & 958 & 0.69 & 270 & 0.28 & 3 & Uncertain \\
135 & 2MASX J17135221-2321556 & 258.4675 & -23.3654 & 0.0285 & 690 & 0.50 & -300 & 0.31 & 3 &  \\
137 & 2MASX J17135847-2251566 & 258.4937 & -22.8657 & 0.0297 & 1307 & 0.95 & 60 & 0.06 & 5 &  \\
138 & 2MASX J17140172-2313026 & 258.5072 & -23.2174 & 0.0284 & 834 & 0.60 & -330 & 0.35 & 3 &  \\
139 & 2MASX J17140971-2331536 & 258.5405 & -23.5316 & 0.0293 & 901 & 0.65 & -60 & 0.06 & 5 & Uncertain \\
145 & 2MASX J17150845-2354140 & 258.7852 & -23.9039 & 0.0303 & 1736 & 1.26 & 240 & 0.25 & 3 & \citet{galdeano22} reports $z=0.084$ \\
146 & 2MASX J17150925-2341020 & 258.7885 & -23.6839 & 0.0324 & 1478 & 1.07 & 869 & 0.91 & 5 & Uncertain \\
149 & 2MASX J17154530-2340044 & 258.9388 & -23.6679 & 0.0284 & 1733 & 1.25 & -330 & 0.35 & 3 & Uncertain \\
151 & 2MASX J17165090-2324074 & 259.2121 & -23.4021 & 0.0286 & 2150 & 1.56 & -270 & 0.28 &  &  \\
152 & 2MASX J17181000-2313584 & 259.5417 & -23.2329 & 0.0262 & 2812 & 2.04 & -989 & 1.04 &  &  \\
\hline
* & 2MASX J17123594-2310058 & 258.1497 & -23.1683 & nan & 435 & 0.32 & nan & nan & 5 & Jellyfish \\
** & VVVX J171253.23-235041.0 & 258.2218 & -23.8448 & nan & 1035 & 0.75 & nan & nan & 0 & Uncertain \\
\multirow{2}*{***} & 2MASS J17132720-2240478 & \multirow{2}*{258.3639} & \multirow{2}*{-22.6798} & \multirow{2}*{nan} & \multirow{2}*{1552} & \multirow{2}*{1.12} & \multirow{2}*{nan} & \multirow{2}*{nan} &  & \multirow{2}*{Interacting galaxies?} \\
& 2MASX J17132811-2240546 & &  &  &  &  &  &  & &  \\
  \hline
  \end{tabular}
  \tablefoot{The top part of the table refers to the spectroscopically confirmed cluster members from \citet{durret15}, while the three entries at the bottom indicate three radio sources likely associated with cluster members. The identification number (ID) and redshift ($z$) are from \citet{durret15} while the morphological class (Mtype) is from \citet{galdeano22}. The projected distance ($r$) and velocity offset ($\Delta v$) are computed from the BCG ($\rm{RA}_{\rm BCG} = 258.1155\deg$; $\rm{DEC}_{\rm BCG} = -23.3698\deg$, and $z_{\rm BCG} = 0.0295$). The normalized distance ($r/\rfive \approx r/(0.65\rtwo)$) and velocity offset ($|\Delta v|/\sigma$) are computed using the values reported in Table~\ref{tab:properties}.}
\end{table*}

\end{appendix}

\end{document}